\documentstyle[12pt]{article}
\begin{document}

 \def\figcap{\section*{Figure Captions\markboth
        {FIGURECAPTIONS}{FIGURECAPTIONS}}\list
        {Figure \arabic{enumi}:\hfill}{\settowidth\labelwidth{Figure
 999:}
        \leftmargin\labelwidth
        \advance\leftmargin\labelsep\usecounter{enumi}}}
 \let\endfigcap\endlist \relax

\begin{titlepage}

\begin{center}
{\bf Budker Institute of Nuclear Physics}
\end{center}

\vspace{1cm}

\begin{flushright}
BINP 94-47\\
May 1994
\end{flushright}

\bigskip

\begin{center}
{\bf NUCLEAR MAGNETIC QUADRUPOLE MOMENTS}
\end{center}

\begin{center}
{\bf IN SINGLE-PARTICLE APPROXIMATION}
\end{center}

\begin{center}
V.F. Dmitriev\footnote{e-mail address: dmitriev@inp.nsk.su},
I.B. Khriplovich\footnote{e-mail address: khriplovich@inp.nsk.su}
and V.B. Telitsin\footnote{e-mail address: telitsin@inp.nsk.su}
\end{center}

\begin{center}
Budker Institute of Nuclear Physics, 630090 Novosibirsk,
Russia
\end{center}

\bigskip

\begin{abstract}
Static magnetic quadrupole moment of a nucleus, induced by T- and P-odd
nucleon-nucleon interaction, is investigated in the single-particle
approximation. Models are considered allowing for analytical solution. The
problem is also treated numerically in a Woods-Saxon potential with
spin-orbit interaction. The stability of results is discussed.
\end{abstract}

\vspace{7cm}

\end{titlepage}
1. Magnetic quadrupole moment is a static characteristic of a quantum
system which is forbidden by P- and T-invariance. Nuclear magnetic
quadrupole moment (NMQM) can be induced both by the nucleon electric
dipole moment\cite{kh} and by P- and T-odd nuclear forces\cite{sfk}. The
interest to NMQM is due to the experimental searches for P- and T-odd
effects in atoms and molecules (see, e.g., book\cite{khr}).

The manifestation of electric dipole moment (EDM), which also violates P-
and T-invariance, in atomic and molecular phenomena is strongly hampered
by the electrostatic screening. In a stationary system of nonrelativistic
pointlike particles interacting via Coulomb forces such a screening of
average electric field acting on any paricle is complete. Therefore, in
such a system for a particle EDM  there is nothing to interact with,
which means that this EDM just cannot be observed\cite{pr,gl}.

The nuclear dipole moment becomes observable however due to the finite
size of a nucleus, more exactly due to different distribution of its
charge and EDM\cite{sch}. One more way to transfer nuclear P- and
T-violation to an atom or molecule is via NMQM\cite{kh}. It demands of
course nuclei with spin $I>1/2$. Besides, the NMQM induces P- and T-odd
effects only in atoms (molecules) with unpaired electron angular momenta
since it interacts directly with magnetic field of the electrons.
However, when operative, the NMQM is much more effective for circumventing
the electrostatic shielding in atoms and molecules\cite{kh,sfk,khr}.

It has been shown in Ref.\cite{sfk} that the NMQM induced by the
P- and T-odd internucleon interaction can be much larger than that due to
the nucleon EDM. In that paper the quadrupole moments, induced by that
interaction, were evaluated in a simple model where the profile of nuclear
density was assumed to coincide with that of nuclear potential. In the
present article we calculate NMQM within a more accurate approach. Namely,
we use a realistic description of the nuclear density; the nucleon wave
functions and Green's functions are obtained with Woods-Saxon potential
which includes the spin-orbit interaction. We include also the
contribution of the current generated by the spin-orbit interaction;
contrary to naive expectations, this contribution does exist for an outer
neutron, but does not for an outer proton. One more model admitting a
closed analytical solution is considered, that of the oscillator
potential. We restrict throughout the present paper to the single-particle
approximation, that of a valence nucleon above a spherically-symmetric
core.

This approach was recently used by us\cite{dkt} for treatment of nuclear
anapole moments, P-odd, but T-even characteristic.

2. Let us begin with discussing the T-and P-odd nucleon-nucleon potential.
In the local limit and to first order in the nucleon velocities $p/m$ it
can be written as follows (see, e.g., book\cite{khr})
\begin{eqnarray}\label{tb} \nonumber
W_2  =  \frac{G}{\sqrt{2}}\frac{1}{2m} \sum_{a,b} \left( ( \xi_{ab}
\vec{\sigma}_a   - \xi_{ba} \vec{\sigma}_b ) \cdot \vec{\nabla}
\delta ( \vec{r}_a - \vec{r}_b )\right.\\
     +  \xi^\prime_{ab}\; [\vec{\sigma}_a
\times \vec{\sigma}_b] \cdot \{ (\vec{p}_a - \vec{p}_b),
\delta ( \vec{r}_a - \vec{r}_b )\} \left)\right.
\end{eqnarray}
where the notation $\{\;\;\;,\;\;\;\}$ means anticommutator. The
dimensionless constants $\xi$ characterize the magnitude of the
interaction in units of the Fermi weak interaction constant $G =
10^{-5}/m^2$ and are supplied with subscripts in order to
distinguish between protons and neutrons.

After averaging this expression over the core nucleons we obtain the P-
and T-odd mean field potential for an outer nucleon
\begin{equation} \label{tvpot}
W(\vec{r}) = \frac{G}{\sqrt{2}}\frac{\xi_a}{2m} \vec{\sigma}\cdot
\vec{\nabla}\rho(r)\ .
\end{equation}
Here $\rho (r)$ is the density of the core nucleons normalized by the
condition $\int d \vec{r} \rho (r) = A$ $(A \gg 1)$;
$$\xi_a = \xi_{ap} \frac{Z}{A} + \xi_{an} \frac{N}{A}\ ,$$
the subscript $a$ takes the values $p$ and $n$ for an outer proton and
neutron, respectively.

Let us note that, as distinct from the case of the P-odd, T-even
interaction, no contact current is generated here in the single-particle
approximation, even if one starts from the two-body interaction (\ref{tb}).
Indeed, it is only the last term in (\ref{tb}), dependent on
$\xi^{\prime}_{ab}$, which contributes to the contact current operator
\begin{equation}
\hat{\vec{j}_c} = \frac{i}{2} \sum_{a} \{ [W_2, e_a \vec{r}_a],
\delta ( \vec{r} - \vec{r}_a )\}.
\end{equation}
However, even this contribution vanishes obviously after averaging over
the core nucleons.

Now, the correction  $\delta \Psi$ to the valence nucleon wave function
generated by the interaction (\ref{tvpot}) is a solution of the equation
\begin{equation} \label{schrod}
(\hat{H}_0 - E)\delta\Psi(\vec{r}) = -W(\vec{r})\Psi(\vec{r}),
\end{equation}
where $\hat{H}_0$  and
$\Psi(\vec{r}) $ are the unperturbed mean field Hamiltonian and the
unperturbed nucleon wave function.  To begin with, let us discuss a
simple model where the profiles of the nuclear density and the central
mean field potential coincide, and the spin-orbit
potential is absent\cite{sfk}
$$ \rho(r) =
-\frac{\rho_0}{U_0}U(r).$$
Eq.(\ref{schrod}) transforms here as follows:
\begin{equation} \label{schrodm}
(\hat{H}_0 - E)\delta \Psi(\vec{r}) = -\imath \frac{G}{\sqrt{2}}
\frac{\xi}{2m}\frac{\rho_0}{U_0}  [\hat{H_0}\,
,\,\vec{\sigma}\cdot\hat{\vec{p}}]\Psi(\vec{r}),
\end{equation}
which gives
\begin{equation}\label{corm}
\delta\Psi(\vec{r}) = -
\imath\frac{G}{\sqrt{2}}\frac{\xi}{2m} \frac{\rho_0}{U_0}
(\vec{\sigma}\cdot \hat{\vec{p}})\Psi(\vec{r}) = -
 \frac{G}{\sqrt{2}}\frac{\xi}{2m}\frac{\rho_0}{U_0} (\vec{\sigma}
\cdot \vec{n}) \Omega_{Ilm}(\vec{n})\left( \frac{dR(r)}{dr} +
\frac{1+K}{r} R(r)\right).
\end{equation}
Here $\;\Omega_{Ilm}$ is a spherical spinor, $\;R(r)$ is the
unperturbed radial wave function of a nucleon, and $\;K= (l-I)(2I+1)$.

Even in a more general case, beyond this model, it is convenient to define
the correction $\delta R(r)$ to radial wave function by the following
relation:
\begin{equation} \label{deltaR}
\delta\Psi(\vec{r}) = - \frac{G}{\sqrt{2}}\,\xi\,\rho_0\,
(\vec{\sigma}\cdot \vec{n}) \Omega_{Ilm}(\vec{n}) \delta R(r).
\end{equation}
The correction $\delta R(r)$ can be calculated using two
independent solutions of the radial Schr\"odinger equation $u_1(r)$
and $u_2(r)$, regular at the origin and at the infinity,
respectively.  These solutions are normalized to the unit Wronskian:
$$u_1\,\frac{du_2}{dr} - \frac{du_1}{dr}\,u_2 = 1.$$
This correction is
\begin{equation} \label{deltarad}
\delta R(r) = - \frac{u_{1,Il'}}{r}
\int_r^{\infty} dr'\, u_{2,Il'}(r')\frac{df(r')}{dr'}u_0(r') -
\frac{u_{2,Il'}}{r} \int_0^r dr'\,
u_{1,Il'}(r')\frac{df(r')}{dr'}u_0(r') ,
\end{equation}
where
$l'=2I-l$, $u_0(r) = rR(r)$ and $f(r)$ is a density profile $f(r)=
\rho(r)/\rho_0$.

The magnetic quadrupole moment operator $\hat{M}_{ij}$ is defined by
analogy with the electric quadrupole one $\hat{Q}_{ij}$, via the
interaction with the corresponding field gradient:
\begin{eqnarray}
\hat{H}_{Q}=-\frac{1}{6}\hat{Q}_{ij}\nabla_iE_j, \\ \nonumber
\hat{H}_{M}=-\frac{1}{6}\hat{M}_{ij}\nabla_iB_j.
\label{eq:def}
\end{eqnarray}
The symmetric tensor $\hat{M}_{ij}$ is related in the following way to the
current density $\hat{J}_n$:
\begin{equation}
\hat{M}_{ij} = \int d\vec{r} (r_i \epsilon_{jmn} + r_j \epsilon_{imn})
r_m \hat{J}_n.
\end{equation}
For a valence nucleon this operator can be presented as\cite{sfk,khr}
\begin{equation} \label{mqms}
M_{ij} = \frac{e}{2m}\left(3\mu (r_i \sigma_j + r_j \sigma_i -
\frac{2}{3}\delta_{ij}(\vec{\sigma} \cdot \vec{r})) +
 2q(r_i l_j + r_j l_i)\right),
\end{equation}
where $\mu$ is the nucleon magnetic moment, and $q$ is equal 1 for
a proton and 0 for neutron.

With the usual definition
$$ M = \langle I m=I| M_{zz}|I m=I \rangle$$
one obtains after taking expectation value over angular variables
\begin{equation} \label{mqv}
M = \frac{G}{\sqrt{2}}\,\xi\,\rho_0\,\frac{e}{m}(\mu -q)
\frac{2I-1}{I+1}(\delta R|r|R).
\end{equation}
The radial matrix element here is
\begin{equation} \label{rme}
(\delta R|r|R) = \int_0^{\infty} r^2dr\,\delta R(r)rR(r).
\end{equation}

For the simple model described above (\ref{corm}) the matrix element
can be calculated analytically with the following result
\begin{equation}  \label{r0}
(\delta R|r|R) = \frac{K-1/2}{2mU_0}.
\end{equation}

One more model allowing for an exact analytical result for MQM
is that of the oscillator potential. Here it is
convenient to start from expression (\ref{rme}) for the
matrix element. Separating the tensor structure in (ref{tvpot}),
(\ref{mqms}) one obtains
\begin{equation} \label{hme}
(\delta R|r|R)
= -\frac{1}{4m}\sum_n \frac{\langle
0|f'(\vec{\sigma}\cdot\vec{n})|n\rangle\langle
n|\vec{\sigma}\cdot\vec{r}|0\rangle + \langle 0|\vec{\sigma}
\cdot\vec{r}|n\rangle\langle n|f'(\vec{\sigma}\cdot\vec{n})
|0\rangle}{E_0 - E_n}.
\end{equation}
For a harmonic oscillator
$$(\vec{\sigma}\cdot\vec{r}) = \frac{
\imath}{m\omega^2}[(\vec{\sigma}\cdot\vec{p}),H].$$
Substituting this identity into (\ref{hme}) and using the completeness
relation, we find
$$(\delta R|r|R)= -
\frac{\imath}{4m^2\omega^2}\langle 0|
[f'(\vec{\sigma}\cdot\vec{n}),\vec{\sigma}
\cdot\vec{p}]|0\rangle.$$
Taking again expectation value over angular variables, we obtain
\begin{equation} \label{oscme}
(\delta R|r|R)= \frac{1}{4m\omega^2}\langle f''
 -\frac{2K}{r}f'\rangle.
 \end{equation}

3. Expression (\ref{mqms}) for the MQM corresponds to the
contribution of the convection and the spin electromagnetic
current densities. In this section we discuss one more contribution to MQM,
that originating from the momentum dependence of the spin-orbit two-nucleon
forces. In the single-particle approximation this current density
is\cite{dkt}
\begin{equation} \label{jpls}
\vec{j}^p_{ls} =
eU^{pn}_{ls}\rho_0\frac{N}{A}\frac{df(r)}{dr}\,\vec{\sigma}\times\vec{n}
\end{equation}
for a valence proton and
\begin{equation} \label{jnls}
\vec j_{ls}^n = -eU_{ls}^{pn}\frac{Z}{A}\, \rho_0f(r) \,
\vec{\nabla}\times \left( \psi^{\dagger
}(\vec{r})\vec{\sigma}\psi(\vec{r})\right)
\end{equation}
for a valence neutron. Here $U^{pn}_{ls}$ is the constant entering two-body
proton-neutron spin-orbit treated in the contact limit:
\begin{equation}
U_{ls} = \frac{1}{2} \sum_{pn} U^{pn}_{ls}\, (\vec{p}_p - \vec{p}_n)
\cdot (\vec{\sigma}_p + \vec{\sigma}_n) \times \vec{\nabla} \delta
(\vec{r}_p - \vec{r}_n).
\end{equation}
The proton-proton spin-orbit interaction does not
contribute to the current density in the zero-range limit we use.

The direct calculation shows that, contrary to naive expectations, the
proton spin-orbit current (\ref{jpls}) does not contribute to the static
NMQM. However, for an outer neutron the corresponding correction does not
vanish. It equals
\begin{equation} \label{mls}
M^n_{ls} =
-2\frac{G}{\sqrt{2}}\,\xi\,\rho_0^2\,
eU^{np}_{ls}\frac{Z}{A}\,\frac{2I-1}{I+1} \, (\delta R|rf(r)|R).
\end{equation}

4. We are ready now for a more realistic single-particle calculation. This
numerical treatment is based on the Woods-Saxon potential including spin-orbit
interaction and on a realistic description of nuclear density.
The profiles of both density and the central part of nuclear potential are
described by a Fermi-type function
\begin{equation}
f(r) ={ 1 \over{1 + exp({{r-R}\over{a}})}},
\label{fermi}
\end{equation}
The total single-particle potential $U(\vec{r})$ is chosen
in a standard Woods-Saxon form
\begin{equation}
U(\vec{r}) = U_{0}f(r) +
U_{ls}\frac{1}{r} \frac{df(r)}{dr}(\vec{l}\vec{\sigma }) + U_{C}(r),
\label{wspot}
\end{equation}
where $U_{C}(r)$ is the Coulomb potential of
a uniformly charged sphere.

We use the values of the density parameters from book\cite{bm}:
\begin{equation}
R = 1.11\,A^{1/3}\;fm;\;\;\; a = 0.54\; fm\;\;\; \rho_0 = 0.17\; fm^{-3}.
\end{equation}
The Woods-Saxon potential is parametrized as in\cite{chp}:
\begin{eqnarray} \nonumber \label{chpls}
R = R_{ls} = 1.24A^{1/3}\; fm,\;\;\; a = a_{ls} = 0.63\; fm, \\
U_0 = (-53.3 \pm 33.6\,\frac{N-Z}{A})\; MeV,\;\;\; U_{ls} = -0.263\,(1+2\,
\frac{N-Z}{A})\,U_0
\end{eqnarray}
The spin-orbit interaction constant as fitted in Ref.\cite{sad74} is:
\begin{equation}
U^{pn}_{ls} = 134.3\; MeV\cdot fm^5.
\end{equation}

The correction $\delta R(r)$ calculated in this way is plotted in Fig.1
together with the model function (\ref{corm}). Obviously, the latter is
a reasonably good approximation to the correction $\delta R$, as
calculated numerically in the more realistic approach.

Our results are conveniently presented in terms of the
dimensionless constant $\tau$ related to the NMQM as follows
\begin{equation} \label{mred}
M = \frac{G}{\sqrt{2}}\xi \, \frac{2I-1}{2I+2}\,e\,\tau.
\end{equation}
This constant itself consists in general of two contributions:
\begin{eqnarray} \nonumber
\tau        & = & \tau_0 +\tau_{ls},               \\
\tau_0      & = & \frac{\rho_0}{m_p}(\mu - q)\, 2(\delta R|r|R), \\
\tau_{ls}^n & = & -4U_{ls}^{np} \frac{Z}{A} \rho_0^2 (\delta R|rf(r)|R),\\
\tau_{ls}^p & = & 0.
\nonumber
\end{eqnarray}

\newpage
The values of $\tau$ calculated for two neighbouring nuclei, with odd Z
and odd N respectively, are presented in Table 1.

5. In conclusion let us compare the results obtained for the NMQM,
generated by P- and T-odd potential, with those for the nuclear anapole
moment (AM), generated by P-violating, but T-even potential. In
particular, we wish to compare the stability of nuclear single-particle
calculations for those two moments, T-odd and T-even.

For the sake of comparison with the constant $\tau$ calculated here, it is
convenient to delete from the dimensionless AM characteristic $\kappa$
the fine structure constant $\alpha$ (related to the electromagnetic
interaction of an atomic electron with nuclear AM) and the P-odd
nucleon-nucleon constant $g$ (the T-even analogue of the constant $\xi$
used here). The typical value of this AM characteristic is\cite{fks,khr,dkt}
\begin{equation}
\frac{\kappa}{\alpha g}\simeq \frac{\mu}{m r_0}\;A^{2/3},
\end{equation}
where $r_0=1.2 fm$.

As to $\tau$, its typical value is
\begin{equation}
\tau\simeq{\mu K}{4m^2 U_0 r_0^3}.
\end{equation}

The ratio of those two factors, i.e., of the T-odd effect to T-even one,
\begin{equation}
\frac{K}{4m U_0 r_0^2}\;A^{-2/3}\simeq 0.15\,K\;A^{-2/3}
\end{equation}
is very small. The origin of the AM enhancement $\sim A^{2/3}$ can be
traced back to the AM dependence on the geometrical cross-section of
nucleus, this is a bulk nucleus effect\cite{fks,khr,dkt}. As to the NMQM,
its magnitude depends completely on the nuclear boundary (see
eqs.(\ref{tvpot}), (\ref{oscme})).

It results not only in the relative suppression of the T-odd effect. The
value of NMQM is more sensitive to the details of the nuclear model than
that of AM, it is less stable.

However reliable theoretical predictions both for AM and
NMQM can be obtained only when the single-particle calculations will be
supplemented by a serious treatment of many-body effects.

\bigskip
This investigation was financially supported by the Russian Foundation for
Fundamental Research.

\pagebreak

\newpage

  \begin{table}[t]
  \begin{center}
  \begin{tabular}{ccccc}
 Nucleus      &   &$\rho=-U\rho_0/U_0$& Harmonic oscillator & Woods-Saxon \\
              &   &                   &                     &         \\
 $^{133}$Cs ($1g_{7/2}^p$) &$\tau = \tau_0$& 0.16  &  0.26 & 0.18 \\
                           &               &       &       &      \\
 $^{137}$Ba ($2d_{3/2}^n)$ &$\tau_0$       &-0.09  & -0.17 & -0.12 \\
                           &$\tau_{ls}$    &-0.02  &       & -0.02 \\
                           &$\tau$         &-0.11  &       & -0.14 \\
        \end{tabular}
  \end{center}
  \caption{The dimensionless MQM as calculated in different approaches}
  \label{tab:r}
  \end{table}

\newpage

\section*{Figure Caption}
{\bf Figure 1.}\ \ \  The correction $\delta R(r)$ for $^{137}Cs$.
Dashed line is the
model function (\ref{corm}). Full line is the $\delta R$ in Woods-Saxon
potential

\end{document}